# Acceleration of Free Electrons by Photonic Time-Crystals

Lior Bar-Hillel[1], Alexey Gorlach[2], Ido Kaminer[1] and Mordechai Segev[1,2]

1. *Department of Electrical and Computer Engineering, Technion, Haifa 32000, Israel*
2. *Physics Department, Technion, Haifa 32000, Israel*

## Abstract

We study the quantum interaction between free electrons and photons in a time-varying media, and find that periodic modulation exponentially amplifies electron-photon coupling within momentum gaps, enabling arbitrarily large momentum transfer. By preparing the light in a two-mode squeezed vacuum state, the electron momentum grows faster than its spectral spreading, establishing time-modulated photonic media as a platform for accelerating free-electrons and shaping their quantum state.

Time has been playing a unique role in physics: it is the only unidifectional dimension and seems to be largely uncontrollable. Recently, the concept of time as a new degree of freedom has gained significant attention, in the context of waves in time-dependent environments. The most fundamental example is a wave traveling through a medium undergoing a sudden change in its electromagnetic (EM) properties, e.g., refractive index. This "temporal boundary" forces the wave to split into forward-propagating and backward-propagating components, known as time-refraction and time-reflection. While a temporal boundary may seem analogous to a spatial interface, it is governed by a fundamentally different set of conservation laws. Namely, at a spatial boundary, energy is conserved while momentum is not; conversely, at a temporal boundary, space remains homogeneous while time-translation symmetry is broken. This results in the conservation of momentum rather than energy. Consequently, the wave-vector of the light remains unchanged across the temporal boundary, while its frequency shifts to accommodate the new refractive index [1]. When the refractive index experiences time-interfaces periodically at sub-cycle rates and order-unity magnitude, a Photonic Time-Crystal (PTC) forms, giving rise to Floquet states. These states form a bandstructure that mirrors spatial crystals, but with crucial distinctions: the bands are defined in momentum and are separated by momentum gaps (k-gaps) [2,3]. The dynamics of light within these gaps is the hallmark of PTCs. Modes with wavenumbers inside momentum gaps have complex frequencies, appearing in conjugate pairs, growing or decaying exponentially over time. This unique amplification mechanism arises directly from breaking time-translation symmetry, where the modulation transfers energy to the EM field in a non-resonant manner. PTCs can have major impact on light-matter interactions: they are predicted to exponentially increase the photon number for modes residing within their k-gaps and induce exponentially growing squeezed states from vacuum fluctuations. Likewise, PTCs were predicted to allow control over the emission rates of atoms [4,5], much like how spatial photonic crystals can inhibit spontaneous emission [6]. Another example is the emission of light by free electrons propagating within a PTC, even below the Cherenkov threshold [7,8]. However, these studies treated the electron momentum as fixed, and therefore did not address the inelastic recoil accompanying coherent electron-light scattering. In reality, free electrons can absorb and emit discrete quanta of an optical near field, acquiring recoil in the process — a phenomenon prominently realized in photon-induced near-field electron microscopy (PINEM) [10]. Over the years, PINEM has evolved from a spectroscopic and imaging tool into a broader platform for coherent control of free-electron wavefunctions and their interaction with quantum light [11-26]. The interaction is strongest under phase-matching, where the electron velocity is synchronized with the optical phase velocity. Recent experiments have demonstrated this regime across a wide range of systems, including metallic near fields, photonic cavities, whispering-gallery modes, dielectric laser accelerators, integrated photonic waveguides [11,14–20,26]. These advances established free electrons as ultrafast nanoscale probes of photonic states, with femtosecond temporal resolution, subwavelength spatial access, and sensitivity to quantum features [17–24,27-30]. Yet, thus far PINEM and free-electron quantum-optics theories were developed exclusively for stationary media. Thus, exploring PINEM interactions within time-varying media opens an exciting new frontier: understanding how coherent inelastic electron-light scattering is modified when the photonic environment itself varies in time, and conversely, whether the electron spectrum can serve as a direct probe of the ultrafast temporal dynamics of such media.

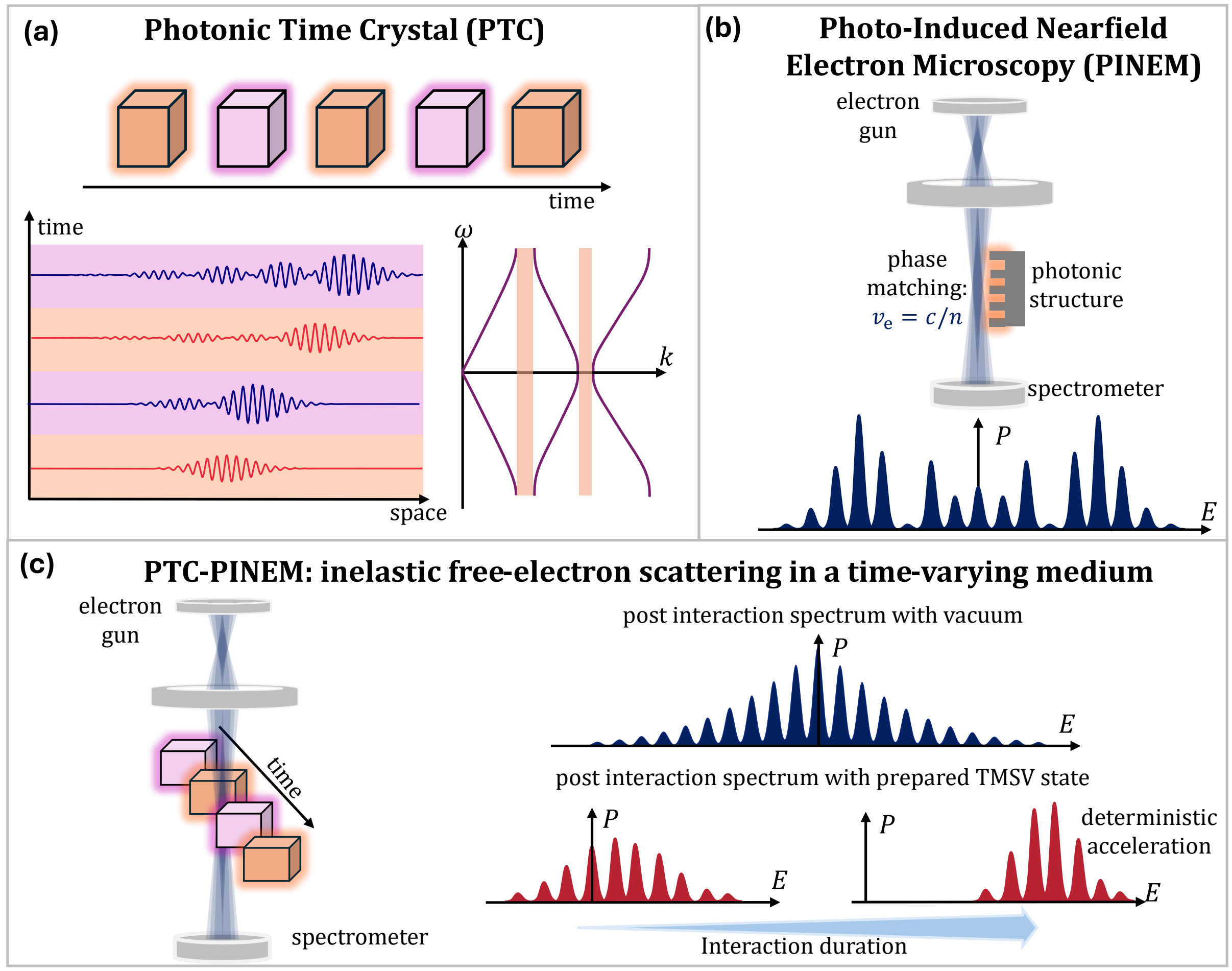

**Fig. 1: Inelastic free-electron scattering in a photonic time-crystal. (a)** Photonic time-crystal (PTC) formed by refractive index that is periodically modulated in time. The resulting Floquet states form a band structure with momentum gaps where the quasifrequencies are complex and the optical field can be exponentially amplified. **(b)** Conventional PINEM: a free electron interacts inelastically with the optical near field, absorbing and emitting discrete quanta of energy and producing a ladder of energies in the electron spectrum. **(c)** PTC–PINEM setting: a free electron interacts with the near field of a time-modulated photonic structure. Temporal modulation mixes the forward- and backward-propagating optical modes and enhances the effective inelastic electron-light coupling for modes inside a PTC momentum gap. Consequently, the electron spectrum exhibits enhanced sidebands, modified momentum transfer, and spectral broadening. This identifies PINEM as a natural probe of sub-cycle PTC dynamics, while simultaneously establishing PTCs as a platform for shaping free-electron quantum states.

Here, we investigate the quantum interaction of free electrons with light within time-varying media, focusing on PTCs. We show that the temporal modulation mixes the forward- and backward-propagating optical modes, renormalizing their coupling to the electron. For modes inside a PTC momentum gap, this coupling grows exponentially with interaction time, resulting in strongly enhanced electron recoil and large broadening of the electron energy spectrum, even away from phase matching, and even when the radiation field is initially in vacuum, thereby establishing free-electron spectroscopy as a natural probe of photonic time-crystals. Remarkably, we find that the net momentum kick imparted to the electron is controlled deterministically by the PTC parameters enabling arbitrarily large, controllable, momentum transfers, with the initial quantum state of light governing the uncertainty accompanying the momentum gain. For example, for a specific initial two-mode squeezed vacuum (TMSV) state, the uncertainty grows only as the square root of the momentum gain. This combination of deterministic, tunable, momentum transfer and quantum uncertainty suppression establishes PTCs as a new platform for the controlled acceleration and deceleration of individual free electrons by light.

**Framework for inelastic free-electron interactions in time-varying media.** Consider a free electron interacting with the near field of a photonic structure whose refractive index varies in time, $n(t)$, Fig. 1c. A relativistic electron propagates along the $z$ axis with velocity $v_e$ in the vicinity of the structure, interacts with its optical near field, and is analyzed after the interaction through its energy spectrum. For simplicity, we restrict the analysis to a one-dimensional model involving two near-field modes with wavevectors $\pm k$ along the electron propagation direction. The full Hamiltonian is

$$H(t) = H_{\mathrm{e}} + H_{\mathrm{ph}}(t) + H_{\mathrm{int}}(t), \tag{1}$$

where $H_{\mathrm{e}}$ is the free-electron Hamiltonian, $H_{\mathrm{ph}}(t)$ is the time-dependent photonic Hamiltonian, and $H_{\mathrm{int}}(t)$ describes the interaction between the electron and the time-varying photonic field. For a relativistic electron, we approximate $H_{\mathrm{e}} = i\hbar v_{\mathrm{e}}\, \partial_z$. The Hamiltonian of the photonic modes in time-varying medium is given by

$$H_{\mathrm{ph}}(t) = \frac{\hbar k c}{n(t)}\left(a_k^\dagger a_k + a_{-k}^\dagger a_{-k} + 1\right) + i\frac{\hbar \dot{n}(t)}{2n(t)}\left(a_k^\dagger a_{-k}^\dagger - a_k a_{-k}\right), \tag{2}$$

where $a_k$ and $a_k^\dagger$ are the annihilation and creation operators of the photonic mode $k$, respectively. The first term describes the time-dependent phase accumulation of the forward- and backward-propagating modes, while the second term captures the mixing and squeezing of these modes induced by the temporal modulation of the medium. Derivation details are provided in the Supplementary Information [Sec. A]. The interaction Hamiltonian is

$$H_{\mathrm{int}}(t) = \frac{\kappa}{\sqrt{n(t)}}\left[\left(a_k + a_{-k}^\dagger\right)e^{ikz} + \left(a_{-k} + a_k^\dagger\right)e^{-ikz}\right], \tag{3}$$

where $z$ is the electron position operator, and $\kappa$ is a coupling strength determined by the interaction geometry, including the electron-structure separation and the properties of the near field. This term describes the coherent exchange of quanta between the electron and the optical field: the electron can absorb from or emit into both the forward- and backward-propagating modes, while the phase factors $e^{\pm ikz}$ encode the corresponding momentum recoil. The explicit appearance of both $k$ and $-k$ channels is crucial in a time-varying medium, where temporal modulation couples the two propagation directions and reshapes the balance between them. The system is initialized in the product state $|\psi\rangle = |p\rangle_{\mathrm{e}}|\psi\rangle_{\mathrm{l}}$, where $|p\rangle_{\mathrm{e}}$ denotes the electron state with momentum $p$ along the $z$ axis, and $|\psi\rangle_{\mathrm{l}}$ is the initial state of the light.

**Photonic time-crystals as an exponential enhancement mechanism.** To analyze the post-interaction electron state, we move to the interaction picture with respect to the reference Hamiltonian $H_{\mathrm{e}} + H_{\mathrm{ph}}(t)$. In this frame, the effect of the time-varying medium is encoded in the evolution of the photonic operators. The annihilation operator of mode $k$ evolves according to a time-dependent Bogoliubov transformation,

$$a_k(t) = u_k(t)\, a_k + v_k(t)\, a_{-k}^\dagger \tag{4}$$

where $u_k(t)$ and $v_k(t)$ are complex coefficients determined by the temporal profile of the refractive index $n(t)$ and the wavenumber $k$. Their derivation, together with verification that the bosonic commutation relations are preserved, is given in the Supplementary Information [Sec. A].

For a PTC, where $n(t) = n(t + T_{\text{PTC}})$, $u_k(t)$ and $v_k(t)$ obey Floquet's theorem and can be written as a periodic function multiplied by $e^{-i\omega_F t}$, where $\omega_F$ is the Floquet quasifrequency set by the PTC bandstructure, Fig. 2a. For modes associated with a momentum gap, $\omega_F$ becomes complex. Consequently, the Bogoliubov coefficients no longer remain bounded in time, but instead contain exponentially growing and decaying contribution simultaniously. This exponential behavior is the hallmark of the PTC and is the origin of the strong interaction enhancement discussed below.

The physical significance of the Bogoliubov coefficients is seen directly in the photon number. For initial vacuum state $|0\rangle_k|0\rangle_{-k}$, the expectation value of the number of photons generated in the two modes is $\langle N(t)\rangle = 2|v_k(t)|^2$. Therefore, for modes inside a momentum gap, the photon number grows exponentially with time, even when the field is initially in vacuum [5].

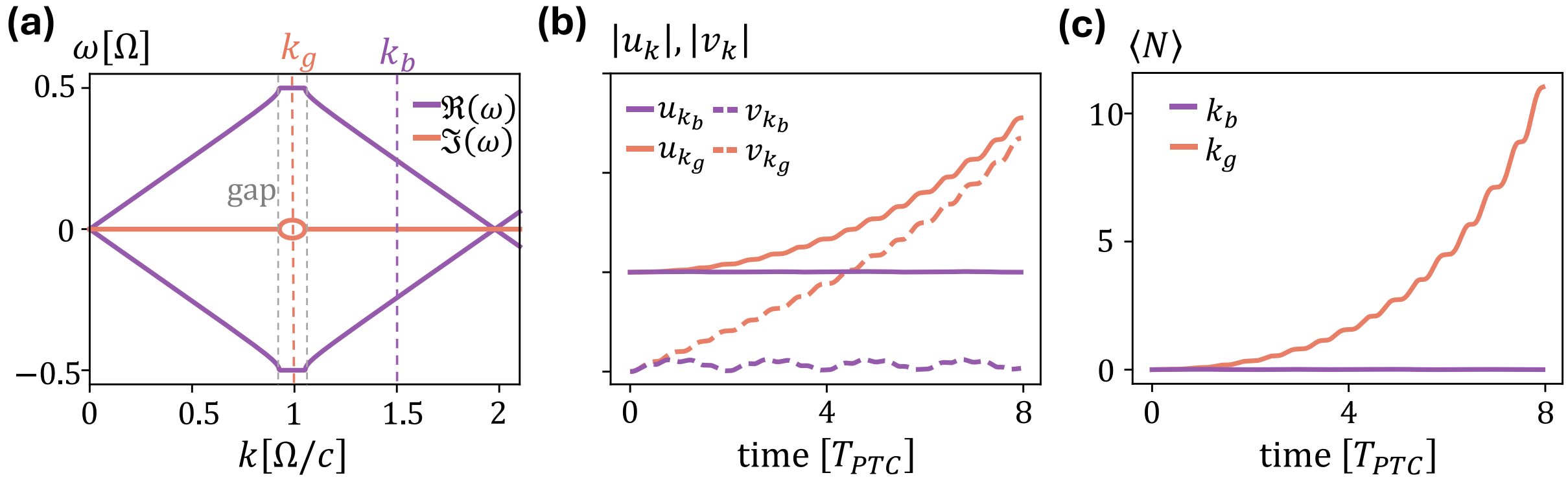

**Fig. 2: Bogoliubov dynamics and exponential amplification in a PTC. (a)** Bandstructure of PTC. Two modes are marked: $k_g$ inside the gap (orange) and $k_b$ outside the gap (violet). **(b)** Time evolution of Bogoliubov coefficients $u_k(t)$ and $v_k(t)$ for the band mode $k_b$ and the gap mode $k_g$. **(c)** Expectation value of the photon number, $\langle N\rangle = 2|v_k(t)|^2$, for the same two representative modes. While the band mode remains weakly populated, the gap mode experiences strong amplification even when the initial state of the field is vacuum.

**Effective PINEM scattering operator in a PTC.** Using Eq. (4), we derive the scattering operator for PINEM interaction between the free electron and the PTC:

$$S = \exp\left[-i\left(g_k a_k b_k^\dagger + g_k^* a_k^\dagger b_k\right)\right] \exp\left[-i\left(g_{-k} a_{-k} b_k + g_{-k}^* a_{-k}^\dagger b_k^\dagger\right)\right], \tag{5}$$

where $b_k = e^{-ikz}$ is the electron momentum ladder operator, satisfying $b_k^\dagger|p\rangle = |p+k\rangle$, and $g_{\pm k}$ are the coupling strengths of the electron and the forward and backward mode. The couplings $g_{\pm k}$ are

$$g_k = \frac{g_k^0}{T}\int_{-\frac{T}{2}}^{\frac{T}{2}}\big(u_k(t)+v_k^*(t)\big)\frac{e^{ikv_\mathrm{e}t}}{\sqrt{n(t)}}dt\,, \qquad g_{-k} = \frac{g_k^0}{T}\int_{-\frac{T}{2}}^{\frac{T}{2}}\big(u_k(t)+v_k^*(t)\big)\frac{e^{-ikv_\mathrm{e}t}}{\sqrt{n(t)}}dt\,, \tag{6}$$

where $T$ is the time of interaction. The constant $g_k^0 = \kappa T$, which can be measured experimentally, represents the coupling strength of a conventional PINEM interaction of duration $T$ in a static medium under phase-matching conditions. The post-interaction state is found by applying the PINEM scattering operator $S$ on the initial state $|\psi_\mathrm{f}\rangle = S|\psi_\mathrm{i}\rangle$.

All the interaction dynamics is contained in the two complex parameters $g_k$ and $g_{-k}$: their magnitudes determine the strength of the inelastic scattering, while their relative phase controls the asymmetry between opposite momentum-transfer channels. Eq. (6) provides direct physical interpretation of the role of temporal modulation. The coupling is determined by the overlap between the electron phase factor $e^{\pm ikv_et}$ and the time-evolving optical mode, encoded in $u_k(t) + v_k^*(t)$. In a static medium, where $u_k(t) = e^{-ikct/n}$and $v_k(t) = 0$, the integral is largest when the optical phase remains synchronized with the electron throughout the interaction. This is the usual phase-matching condition,$v_e = c/n$. Away from this condition, the integrand oscillates rapidly and the net coupling is strongly reduced. PTCs provide a qualitatively different regime: for modes inside a PTC momentum gap, the Bogoliubov coefficients $u_k(t)$ and $v_k(t)$ acquire exponentially growing components. Consequently, the quantity $u_k(t) + v_k^*(t)$ entering Eq. (6) is exponentially amplified, and the effective couplings $g_{\pm k}$ can become much larger than the static PINEM coupling. Physically, the temporal modulation pumps energy into the optical mode, and this enhanced field is then converted into a stronger inelastic recoil of the electron. In this way, the PTC does not merely modify the phase-matching condition; it creates a new enhancement mechanism for electron-light scattering.

The couplings $g_{\pm k}$ can be controlled by tuning the electron velocity, the interaction duration, and the PTC parameters. Their magnitudes determine the broadening of the electron spectrum, while their imbalance and relative phase govern the net momentum transfer. This tunability allows free-electron spectroscopy to probe the amplified dynamics of PTC gap modes, and at the same time facilitates controlled shaping of the electron wavefunction in a temporally structured photonic environment.

**Electron spectrum as a probe of PTCs.** We now turn to the observable that directly connects the time-modulated photonic medium to experiment: the electron spectrum. To isolate the effect of the temporal modulation, first consider the field to be initially in vacuum, $|\psi_l\rangle = |0\rangle_k|0\rangle_{-k}$, where the probability for the electron to acquire momentum change $n\hbar k$ is

$$P_n = \exp[-(|g_k|^2 + |g_{-k}|^2)] \left|\frac{g_{-k}}{g_k}\right|^n I_n(2|g_k g_{-k}|), \tag{7}$$

where $I_n(x)$ is the modified Bessel function of the first kind (see derivation in the Supplementary Material, Sec. B). Equation (7) shows that, even for a vacuum initial state, the electron spectrum can become strongly structured once $g_{\pm k}$ are enhanced by the temporal modulation. In other words, the electron acts as a sensitive readout of the PTC-induced amplification of the optical mode.

As a representative example, consider a sinusoidally modulated permittivity resulting in the refracted index $n(t) = \sqrt{\varepsilon_r + \Delta\varepsilon\cos(\Omega t)}$, where $\varepsilon_r$ is the ambient permittivity, $\Omega$ is the modulation frequency, and $\Delta\varepsilon$ is the modulation amplitude. Fig. 3(a) shows the resultant electron spectrum as a function of momentum transfer and electron velocity. For weak modulation, the spectrum remains concentrated near conventional phase matching, $v_\mathrm{e} \simeq c/\langle n(t)\rangle$, and only a narrow set of sidebands is appreciably populated. As the modulation depth increases, the interaction is dramatically enhanced: the sideband distribution broadens, strong scattering persists away from mean phase matching, and the spectrum develops the characteristic fingerprints of the amplified gap dynamics.

This enhancement is shown in Fig. 3(b), which plots the effective coupling to a representative gap mode, $\left|g_{k_g}\right|^2$ versus the interaction duration and electron velocity. In a static medium, strong interaction is confined to a narrow phase-matched line. In contrast, inside the PTC gap, the coupling grows rapidly with interaction time even under phase-mismatched conditions, because the optical mode itself is exponentially amplified by the temporal modulation. The strongest growth still occurs near mean phase-matching, where the electron remains synchronized with the amplified field for the longest time, but the key point is that the PTC enhancement survives

well beyond this condition. This distinguishes the PTC regime from ordinary PINEM, where phase mismatch quickly suppresses the interaction.

Importantly, the electron spectrum reflects the bandstructure of the PTC. Fig. 3(c) shows the spectrum as a function of the wavenumber. Outside the momentum gap, the interaction is weak, and the spectrum stays narrowly concentrated around the zero-loss line. Inside the gap, by contrast, the exponentially amplified optical mode generates a pronounced ladder of sidebands and strong spectral broadening. The sharp onset of these features at the gap boundaries means that PINEM indicates the position of the momentum gap and serves as a probe of its bandstructure.

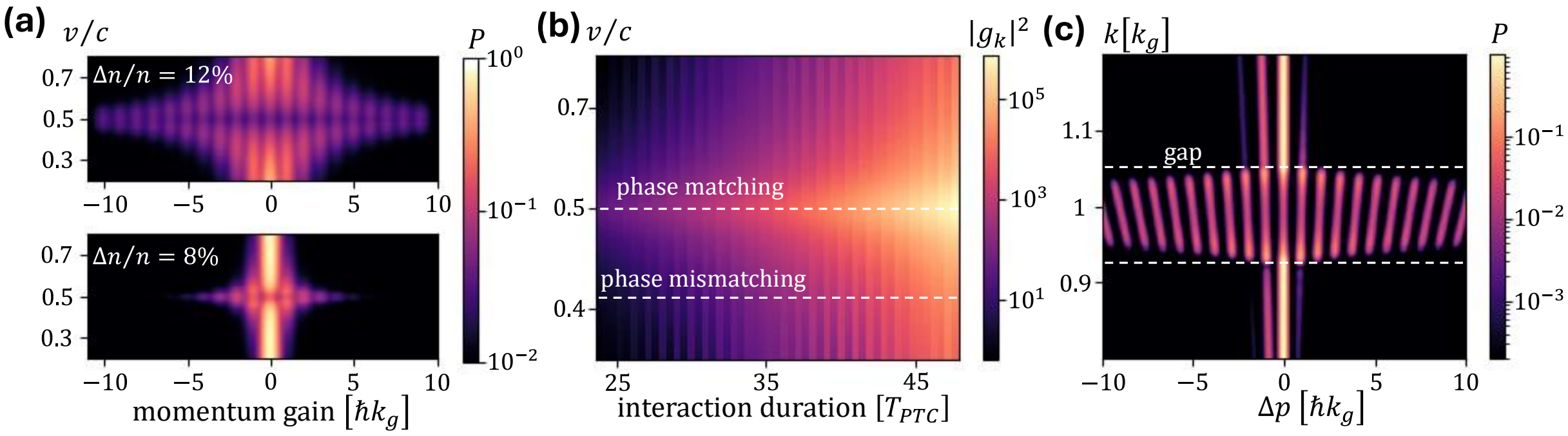

**Fig. 3: Electron spectra as probes of PTC. (a)** Post-interaction electron spectrum $P(\Delta p)$ vs electron velocity $v/c$, for modulation depths $\Delta n/n = 10\%$ and $6\%$. Increasing the modulation depth strongly enhances the inelastic interaction, broadens the sidebands, and extends the scattering well beyond phase-matching conditions. **(b)** Effective coupling strength $\left|g_{k_g}\right|$ for a representative gap mode $k_g$, vs interaction duration and electron velocity $v_e$. The coupling is maximal near mean phase matching, $v_e \simeq c/\langle n(t)\rangle$, but grows strongly with interaction time even away from it due to the exponential amplification of the gap mode. **(c)** Electron spectrum versus photonic wavenumber $k$. Dashed vertical lines mark the boundaries of the PTC momentum gap. Outside the gap, the spectrum remains weak and concentrated near the zero-loss line, whereas inside the gap a pronounced sideband ladder emerges.

**Shaping the electron wavefunction: accelerating individual free electrons.** Beyond using the electron spectrum to probe PTCs, the time-modulated interaction can also be used to shape the electron wavefunction itself. The relevant quantities are the mean momentum transfer and its uncertainty. These distinguish between simple spectral broadening and directional recoil. For the scattering operator derived above, the mean momentum transfer is

$$\langle \Delta p \rangle = \hbar k\left(|g_{-k}|^2 - |g_k|^2\right). \tag{8}$$

This expression reveals two important features. First, the average recoil is determined solely by the couplings $g_{\pm k}$, and therefore by the temporal modulation, interaction duration, and electron velocity; in particular, it is independent of the initial quantum state of the light. Second, the sign and magnitude of the recoil are controlled by the asymmetry between the two scattering channels. By tuning this asymmetry, the electron can be driven toward either net acceleration or net deceleration.

Fig. 4(a) shows the mean momentum transfer versus electron velocity and interaction duration. Regions of positive and negative recoil appear naturally in parameter space, corresponding to acceleration and deceleration. The two marked working points, labeled $A$ and $D$, are representative examples of these opposite regimes. However, a nonzero mean recoil alone is not sufficient for directional control. To obtain a genuinely shifted electron distribution, the average momentum transfer must exceed the momentum uncertainty induced by the interaction. Here, the initial quantum state of the light becomes essential. For generic initial states such as vacuum, coherent light, and thermal light, the momentum uncertainty grows more rapidly than the mean recoil, so the interaction primarily broadens the spectrum rather than shifting it as a whole. A qualitatively different regime arises for suitable nonclassical states of the two optical modes. In particular, for two-mode squeezed vacuum state $|\psi_l\rangle = \exp(r a_k a_{-k} - r^* a_k^\dagger a_{-k}^\dagger)\,|0\rangle_k |0\rangle_{-k}$, with squeezing parameter amplitude $\tanh r = 2|g_{-k}||g_k|/(|g_{-k}|^2 + |g_k|^2)$ and phase $\angle r + \angle g_k + \angle g_{-k} = \pi(2m+1)$ for $m \in \mathbb{Z}$, the momentum uncertainty is

$$\sigma_p = \hbar k \sqrt{\left| |g_{-k}|^2 - |g_k|^2 \right|}. \tag{9}$$

This uncertainty scales as the square root of the average momentum gain $\sigma_p \propto \sqrt{\langle \Delta p \rangle}$. This slower growth of the fluctuations is the central wavefunction-shaping effect: the momentum spread increases much more slowly than the average transfer, so the electron distribution can be displaced rather than merely broadened. An important advantage of the PTC is that it can also assist with the required photonic-state preparation. As discussed above, a PTC naturally generates correlated photon pairs and can create a two-mode squeezed state from vacuum, with controllable squeezing amplitude [5]. Moreover, because the relevant interference depends on the combined phase of $r$, $g_k$, and $g_{-k}$, the interaction duration can be chosen so as to reach the optimal phase relation. In this way, the same temporal modulation that enhances the inelastic coupling can also help prepare the optical state needed to suppress momentum broadening.

Taken together, these results show that PTC-enhanced PINEM is not only a tool for probing time-modulated media but also a framework for engineering the recoil distribution of free electrons. The mean recoil is governed by the asymmetry of the effective couplings, while the fluctuations are set by the quantum state of light. This separation between mean and variance is what enables temporally structured photonic environments to shape free-electron quantum states in ways unavailable in conventional static PINEM.

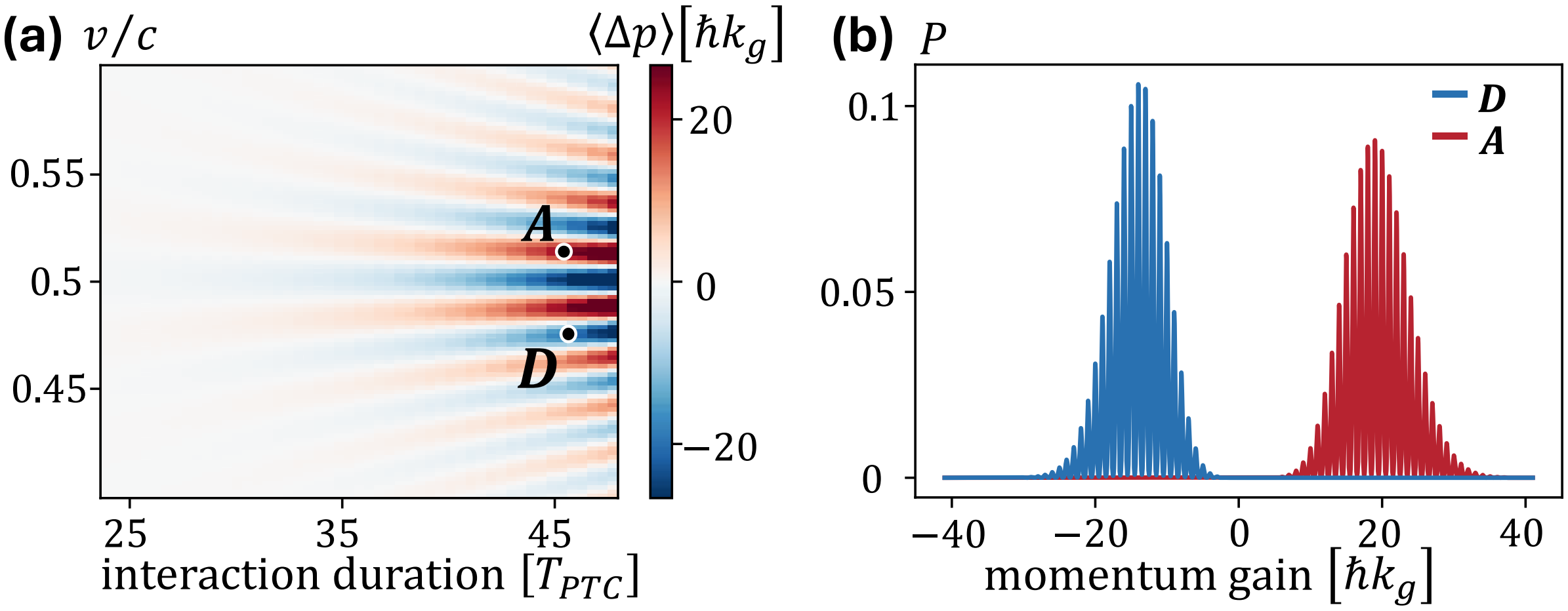

**Fig. 4: Wavefunction shaping through control of the mean recoil and its fluctuations.** **(a)** Mean momentum transfer $\langle \Delta p \rangle$, vs electron velocity $v/c$ and interaction duration. Positive and negative regions correspond to net acceleration and deceleration, respectively. The marked points $A$ and $D$ indicate representative working points used in panel (b). **(b)** Post-interaction electron spectra for the two points marked in (a), obtained for a suitably prepared TMSV optical state. The red curve ($A$) corresponds to a directionally accelerated electron distribution, while the blue curve ($D$) corresponds to deceleration.

**Outlook and Conclusions.** We studied the quantum interaction between free electrons and light within time-varying media, focusing on PTCs. We showed that modes lying inside PTC momentum gaps lead to an exponential enhancement of the electron-light coupling, giving rise to strong inelastic recoil and pronounced post-interaction electron spectra even beyond phase-matching. These results identify PINEM as a natural probe of sub-cycle PTC dynamics and bandstructure. We showed that temporal modulation controls both the mean recoil and its fluctuations, opening a route to shaping free-electron quantum states in temporally structured photonic environments. For suitable nonclassical optical states, this control can suppress spectral broadening relative to the net momentum transfer, enabling directional acceleration. More broadly, our work connects temporal photonics with free-electron quantum optics and suggests a new platform for both probing and engineering ultrafast light-electron interactions.